%% file: elsarticle-template-5-harv.tex
\documentclass[preprint,authoryear,12pt]{elsarticle}
 \usepackage{graphicx}
 \usepackage{epsfig}
\usepackage{adjustbox}
\usepackage{varwidth}
\usepackage{algorithmic}
\usepackage[ruled,vlined]{algorithm2e}
\usepackage{float}
\floatstyle{plaintop}
\restylefloat{table}
\usepackage{amssymb}
 \usepackage{amsthm}
\usepackage[fleqn]{amsmath}

\biboptions{longnamesfirst,comma}

\begin{document}

\begin{frontmatter}

\author{F. Yahyanejad\footnote{The author email addresss: f.yahya2@uic.edu}}
\author{ B. Sadeghi Bigham\footnote{{The author email address: b$\_$sadeghi$\_$b@iasbs.ac.ir}}}
\address{$^1$Department of Computer Science, University of Illinois at Chicago, USA}
\address{$^2$Department of Computer Science, Institute for Advanced Studies in Basic Sciences, Iran}
\title{Greedy Harmony Search Algorithm for the Hop Constrained Connected Facility Location}

\begin{abstract}
We present a simple, robust and efficient harmony search algorithm for the Hop Constrained Connected Facility Location problem (HCConFL). The HCConFL problem is NP-hard that models the design of data-management and telecommunication networks in a manner of reliability. In this paper, we customize harmony search algorithm to solve the HCConFL problem. To arrive to quick, optimal cost of each solution, we use a new greedy approach expanding idea of Kruskal algorithm in our objective function. We also use a new greedy method combined with harmony search to obtain a good approximation in an efficient computational time. The algorithm was evaluated on the standard OR Library benchmarks. Computational results show that with high frequencies the modified harmony search algorithm produces optimal solutions to all benchmarks very quickly. We also solve the problem with another heuristic algorithm including the variable neighborhood search, the tabu search, to evaluate our algorithm.
\end{abstract}

\begin{keyword}
OR in telecommunications \sep Hop constrained Connected Facility Location\sep Harmony search\sep Greedy algorithm
\end{keyword}

\end{frontmatter}

\section{Introduction}
\label{}
 Due to recent growth of telecommunication networks, telecommunication companies have motivated researchers to solutions for network design problems. Such networks are designed to connect a source by intermediate switching devices to subscribers as a network. The intermediate switching devices installed in these networks are so expensive.
  Besides, in the context of reliability, Hob Constraint is used as a limit for the number of intermediate devices used between the source and subscribers. The aim of this paper is to minimize the cost of such networks. Similar problems arise in the design of the communication networks.  \cite{5} have shown that the Fiber-to-the-Curb strategy can be modeled by the connected facility location (ConFL). They have modeled these reliability constraints within the Fiber-to-the-Curb strategy by generalizing the ConFL to the HCConFL.

 The HCConFL problem (Figure 1) is related to two well-known problems: The Facility Location problem and the Steiner tree problem with hop constraints. 

The ConFL problem is HCConFL problem when the hop is infinitive. In ConFL an undirected graph $G=(V, E)$ is given with a dedicated root node $v_0\in V$ and edge costs $c_e\geq 0, \forall e=(u,v)$, Corresponding to the costs of installing a new route between $u$ and $v$. 
Furthermore, a set of facilities $f\subseteq V$ and customer nodes $D\subseteq V$ are given, and also an opening cost $f_i\geq 0$ is assigned to each facility. We try to find a minimum cost tree so that every customer node is assigned to an open facility and also open facilities are connected to the route through a Steiner tree.

\cite{14} first introduced the ConFL. These researchers obtained first  approximation algorithm for this problem. Currently, many research groups have focused on the optimization of the ConFL problem and few heuristic methods are suggested to practical problems. \cite{16} proposed heuristic algorithm for the first time in 2007 by combining tabu search and Neighborhood search. 
Tomazic and Ljubic in 2008 considered the problem without the root and gave the greedy randomized adaptive search procedure in \cite{19}. In 2010, Bardossy and Raghavan gave an algorithm by combining dual ascent approach and neighboring local search to get upper bound and lower bound for the problem \cite{4.5}.

Hop Constrained Steiner Tree problem (HCST): Given an undirected connected graph G=(V, E) and nonnegative weights associated with the edges. Consider a set of essential nodes, a root node, some other non-essential nodes, and also a positive integer $h\leq H$. 
The problem is to find a minimum cost subgraph $T$ of $G$ so that from root to each essential node exists a path $T$ from $v_0\in V$ to each basic node with no more than $H$ intermediate edges (eventually including nodes from $S=(V, Q))$ in \cite{18}.

HCST and Hop Constrained Minimum Spanning tree (HCMST) problems are are very practical in telecommunication network design and network requirements. A recent survey for the HCMST can be found in  \cite{2.5}. Gouveia uses variable redefinition to strengthen a multicommodity flow model for minimum spanning and Steiner trees with hop constraints between a root node and any other node \cite{9}.
 Gouveia in the paper \cite{6} compares the model of multicommodity flow in both directional and non-directional introduced HCST issue in 1998. Then in 1999, Voss presents a mix integer-programming formulation based on Miller-Tucker-Zemlin sub tour elimination constraints and also develops a heuristic algorithm to find the initial solution based on tabu search \cite{20}. 
 \cite{7} proposes a model for HCMST problem based on Miller-Tucker-Zemlin subtour and \cite{5} presents two models based on flow and tree with hop index for the HCMST and HCST in \cite{8}. Santos describes algorithm for the HCST problem in 2010 \cite{18}. In this method, with changing the original graph $G$ into the problem HCST to a layering graph of $G'$, HCST has become change to the Steiner tree and then, the dual ascent algorithm is presented to the Steiner tree problem on a graph $G'$. 
\cite{Botton} study the hop-constrained survivable network design problem with reliable edges and consider two variants of reliable edges when a static problem where the reliability of edges is given, and an upgrading problem where edges can be upgraded to the reliable status at a given cost. 
\cite{Dia} provide integer linear programming formulation for hop constrained network from polyhedral point of view.
Harmony search is a meta-heuristic algorithm, which is inspired by a compositor to compose a piece of music. The harmony search algorithm has been used mostly to solve optimization problems and here we want to utilize this simple and efficient algorithm for a discrete problem (see \cite{21}).

\begin{figure}[h]
\centering
\includegraphics[scale=0.30]{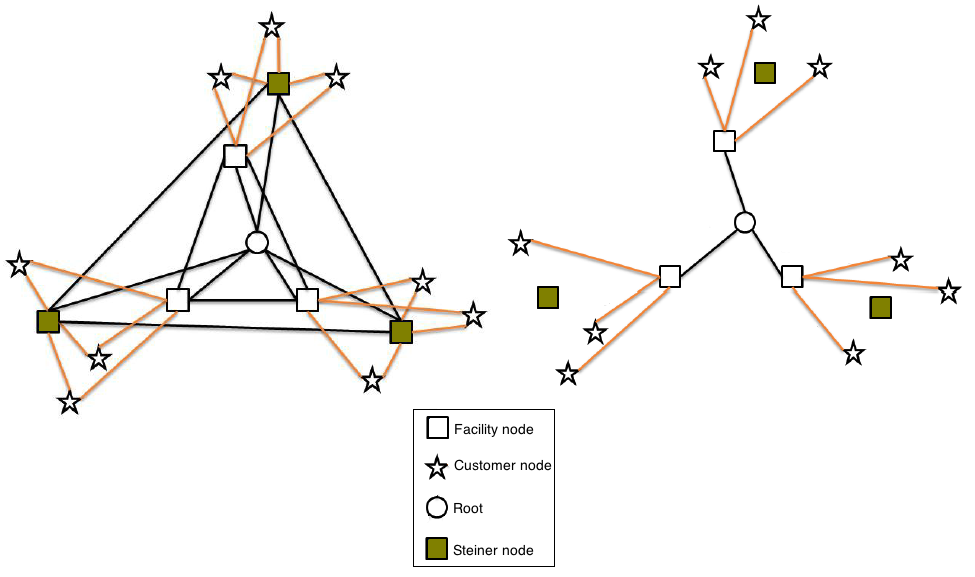}
\caption{1-Constrained Connected Facility Location Example }
\label{figNNfdsds}
\end{figure}

In this paper, we customize harmony search for solving the HCConFL problem, then improving it by combining with a new greedy approach. The paper is organized as follows. In Section 2, the problem is defined. In Section 3, harmony search algorithm is introduced. The customized algorithm details are mentioned in Section 4. The greedy approach for harmony search (modified harmony search) is presented in Section 5. In Section 6 we combine the greedy harmony search with local search. Section 7 is devoted to show the implementation and results and Section 8 provides concluding remarks.
\section{HCConFL formulation}
\label{}
The HCConFL problem can be stated as follows:
Given graph $G = (V, E)$, where $V$ is the set of nodes and $E$ is the set of arcs with cost function $C: E\rightarrow R^+$ for every arc $(i,j)\in E$ which presented with $c_{ij}$. In this graph, the set of nodes $V$ consists of two partitions $\{S, D\}$ that $D$ is set of customers$^\prime$ demands and $S$ is set of core nodes (potential Steiner nodes). 
The set $S$ contains a subset of $F$ called facility set $(F\subseteq S)$. Each facility opening cost is equal to $f_i $ $(i\in F)$ and the node $r ~(r\in F)$ is given as the root node. We determine an optimal location of facilities to fulfill all demands of the customers such that the total cost of establishing the facilities, fulfilling the demands and the cost of Steiner treeÂÂs is minimized as the hop constraint is satisfied.

We should note that only those facilities assigned to the customers pay opening cost. Also, customers in the solution appear as leaves and the out-coming edge of the open facility to the customer does not consider as a hop. 

The problem can be mathematically stated as follows:

\begin{equation}
min \sum_{p=1}^H\sum_{ij \in A_s}c_{ij}x_{ij}^p+\sum_{jk \in A_D}c_{jk}x_{jk}+ \sum_{i \in F} f_iy_i 
\end{equation}
s.t.
\begin{equation}
\sum_{i \in S \backslash {k}:  (i,j) \in A_s}x_{ij}^{p-1} \geq x_{jk}^p \hspace{1.5cm} \forall (j,k) \in A_s,j \neq r,p=2,...,H,
\end{equation}
\begin{equation}
\sum_{(i,j) \in A_s}\sum_{p=1}^Hx_{ij}^{p} \geq y_j \hspace{1.5cm} \forall j \in F / \{r\},
\end{equation}
\begin{equation}
x_{ij}^p =0\hspace{1.5cm}
 (i,j) \in A_S,
 \left\{ 
  \begin{array}{l l}
    1 & \quad \textrm{if i=r,p=2,...,H}\\
    0 & \quad \textrm{p=1,$ i\neq r$}
  \end{array} \right. 
\end{equation}
\begin{equation}
\sum_{(j,k) \in A_D}x_{jk} = 1\hspace{1.5cm} \forall k \in D,
\end{equation}
\begin{equation}
x_{jk} \leq y_{j}\hspace{1.5cm} \forall (j,k) \in A_D,
\end{equation}
\begin{equation}
y_{r} =1,
\end{equation}
\begin{equation}
x_{ij}^p \in\{0,1\}\hspace{1.5cm} \forall (i,j) \in A_S,p=1,...,H,
\end{equation}
\begin{equation}
x_{jk} \in\{0,1\}\hspace{1.5cm} \forall (j,k) \in A_D,
\end{equation}
\begin{equation}
y_i \in\{0,1\}\hspace{1.5cm} \forall i \in F,
\end{equation}
Where,
\begin{equation}
 x_{ij}^p = \left\{ 
  \begin{array}{l l}
    1 & \quad \textrm{if edge $(i,j)\in A_s$ appears in p position to the root.}\\
    0 & \quad \textrm{o.w.}
  \end{array} \right.
\end{equation}
\begin{equation}
 x_{jk} = \left\{ 
  \begin{array}{l l}
    1 & \quad \textrm{If client k is assigned to facility j.}\\
    0 & \quad \textrm{o.w.}
  \end{array} \right.
\end{equation}
\begin{equation}
 y_j = \left\{ 
  \begin{array}{l l}
    1 & \quad \textrm{If facility i is opened.}\\
    0 & \quad \textrm{o.w.}
  \end{array} \right.
\end{equation}

Constraint (2) checks the connection of edges to each other and also because of index $p$, it prevents the creation of cycle. Constraint (3) controls that there is at least one incoming edge to an open facility. Constraint (4) puts some $x_{ij}^p$ equal to zero, where out coming edges of the root can only be placed in the first position. Then, the value of the variables, which their first node is the root, $p\geq 2$ is equaled to zero, and also the source of main edges is zero. Constraint (5) guarantees each customer assigned to one facility and Constraint (6) shows the facility assigning to each customer is opened. Constraint (7) opens the root facility and Constraints (8), (9), and (10) are binary variables indicating the problem. 
\section{Harmony search algorithm}
\label{}
 The harmony search algorithm is a meta-heuristic algorithm, which is inspired by the construction of a new piece of music by a composer. Every piece of music is combined of short pieces of music, rhythm and variant beat of music. A new composer put together a number of samples to create a new piece of music. For example, suppose composer is going to make a piece of music creating $N$ samples. Composer has three options to choose sample $k$: 
 First, using the pieces in the memory that has been already used. Second, using a piece of music in the memory with little changing (pitch) to produce a same sample. Third option is producing of new pieces completely creatively. In the harmony algorithm, each solution vector is composed of $N$ variables as same as the beat of music. To generate a new solution vector (each solution vector is called harmony), we use the same way that a composer uses to create new pieces \cite{21}. \cite{4} formulated three corresponding components for optimization process: using harmony memory, pitch adjustment, and randomization \cite{4}:

1. Each vector is defined as a solution, and we also define a function to evaluate the quality of each solution:
\begin{gather*}
X_i=\{x_i(1),x_i(2),...,x_i(n)\}
\end{gather*}
\begin{gather*}
Minimize \hspace{0.1cm}f(x)
\end{gather*}
\begin{gather*}
s.t. \hspace{0.25cm}x_i\in X_i, i=1,2...,N
\end{gather*}

The main part of harmony search algorithm is harmony memory. The harmony memory size (HMS) is equal to solution vectors with objective function values stored in it to produce new solution vectors.     
\begin{gather*}
HM =
 \begin{pmatrix}
  x_1^1&  x_2^1 & \cdots&  x_N^1& f(x^1) \\
  x_1^2& x_2^2  &\cdots&  x_N^2& f(x^2) \\
  \vdots &  \vdots &  \ddots&  \vdots & \vdots \\
 x_1^{HMS-1}& x_2^{HMS-1}&  \cdots&  x_N^{HMS-1}& f(x^{HMS-1})\\
  x_1^{HMS}& x_2^{HMS}&  \cdots&  x_N^{HMS}& f(x^{HMS})
 \end{pmatrix}
\end{gather*}

At beginning of the algorithm, in initialization, harmony memory fills with distinct and random HMS solution vectors.

2. The process of improvising a new harmony (solution vector): Exactly as mentioned earlier, to produce a new harmonic pattern (solution vector) $X_{new}=(x_1^{new},x_2^{new},...,x_n^{new} )$, we have three options,  choose $x_i^{new}$  among all of $x_i^j$ s in the harmony memory, choose one of the $x_i^j$  and change it a little (pitch), or initialize $x_i^{new}$ with a quite new random value. So we have three parameters, Harmony memory consideration rate ($HMCR$), pitch adjustment rate ($PAR$) and bandwidth ($bw$).

 $PAR$ and $HMCR$ are numbers between 0 and 1 while $PAR$ is always lower than $HMCR$. Each time, a new vector is produced by probably $HMCR$, so one of the $x_i^j$ s are randomly chosen from the memory or it can be chosen with probability of $PAR$ and would be equaled to $x_i^{new}$. Otherwise, it uses a new random value for $x_i^{new}$ with probability of $1-HCMR$. 
 We also consider a parameter that named $bw$. This parameter determines the maximum amount of entries that would be changed. For example, if we want to initialize $x_i^{new}$  by changing the values of an element in memory, it will be as follows:

$X_i=x_i\backslash j(j=rand(0,HMS))\pm rand(0,bw)$ 
\\With this initial description, the algorithm is as follows:

{\bf Step 1}.  Harmony search parameters Initialization ($HMS, PAR, HMCR$ $, bw$),

{\bf Step 2}.  Filling the harmony memory with $HMS$ different random solution vectors.

Algorithm is repeated until the termination condition$\{$

{\bf Step 3}.  Improvising a new harmony (solution vector),

{\bf Step 4}.  Updating the harmony memory,

$x^{New}\in HM\bigwedge x^{Worst}\notin HM,x^{New}\succeq x^{Worst}$,

$\}$

$\succeq: x^{New}$ is valuable than $x^{Worst}$.

{\bf Step 5}.  Consider the best solution in the harmony memory as the final solution.
\section{Customized Harmony search for HCConFL Problem}
\label{}
In Section 3, we described harmony search algorithm. In this section we use this heuristic algorithm to solve the HCConFL problem. Harmony search algorithm is chosen because unlike some heuristic algorithms that are local search algorithms and work by considering neighbors, harmony search is not a local search and because of the random factors contributing to, it can get more variant solutions.
 In addition, it has good power that has never been used in this field. Its special structure makes a good balance between speed and accuracy, speed of convergence, and search dispersion. Also, as we will see in Section 5, using a greedy algorithm, we improve the quality of time and solution of harmony search for this problem. The structure of harmony search algorithm allows widespread of greedy optimization can be combined to it. The harmony search algorithm for HCConFL is presented in Algorithm 1.
 
\bigskip
\begin{algorithm}[H]
\SetKwInOut{Input}{input}\SetKwInOut{Output}{output}
\Input{Objective function $f(x)=\{x_1,x_2,...,x_n \}$,    
              HMCR(ex:0.96), PRA(ex:0), HMS(ex:50).}
\Output{Generate Harmony memory($HM$)with random harmonies
              and add best Harmonies to $HM$.}
\BlankLine
\Begin{
 \While{it $\leq$NUMBER of Iteration}{
 \While{var $\leq$ NUMBER of Variables}{
  rnd=rand(0,1)\;
  \eIf{$HMCR\leq rnd$}{
   Use a random value for $x_{var}^{it}$\;
   }
  { \eIf{PAR$\leq $rnd$\leq$HMCR}{
   Choose a value from all $x_{var}^{it}$  in $HM$\;
  }
  {
   Choose a value from all $x_{var}^{it}$  in $HM$ and
                      adjust it to a close value\;
  }
}
{
   Verify Harmony(in our case:if not a Tree,
               change it to be a Tree with Best objective) and
               Evaluate new harmony and accept if it is
               better than the worse harmony in $HM$\;
  }
 }
 }
}
 \caption{Harmony search algorithm for HCConFL}
\end{algorithm}
\bigskip
\subsection{Defining the objective function:}
As mentioned before, our goal is opening some facilities on some nodes of the graph in such a way that summation on all costs of assigning every customer to exactly one open facility, connecting open facilities via a Steiner tree, and facilities opening cost is minimized regarding the hop constraint. The solution vector is an array with size of number of facilities including zero and one. 
First, we calculate the HCST with minimum cost between open facilities ($x_i=1$ in our generated harmony). We use a new greedy algorithm to find a HCST that connects all selected facilities with minimum cost \cite{200}. The next step is to connect each customer node to the nearest and cheapest facility node which is open. The modified Bellman-ford algorithm calculates the minimum cost paths with hop constraint from the root node to each facility. Richard Bellman and Lester Ford first suggested the graph search algorithm that finds the shortest path between each two nodes with hop constrained  (\cite{2,3}). For calculating the objective function for the HCConFL problem for a specified vector $x$, we have three-step procedure:

Variable $x_i=1$ represents that $i^{th}$ facility is open and $x_i=0$ otherwise; $y_{ij}$ refers to the demand of a customer if $y_{ij}=1$ means customer $j$ is assigned to facility $i$ and $y_{ij}=0$ otherwise;

We use a new efficient greedy algorithm called Not Root Base Insertion (NRBI) to find a Steiner tree Tree with hop constraint to connect all selected facilities $(x_i=1)$. The basic idea to construct HCST and connect open facilities ($Q$ is set of open facilities) is expanding some principles known from algorithms for the MST (\cite{17} and \cite{11} together with a partial solution $G=(\{root\},\emptyset )$ consisting root node $root$. 

The tree is built by insertion of $|Q|-1$ shortest paths to $Q$ while no hop constraints are violated. We start with a partial solution $G=({root},\emptyset)$ that just contains the root. As we mentioned before, $Q$ is the set of all basic nodes. At each step the set $T$ is equal to $Q\backslash G$. Variable $H$ is the maximum number of hops allowed that the root connects to other nodes in the tree. 

The $V_G$ denotes node set of $G$ and $V_T$ denotes the node set of $T$. $d_{uv}$ is the cost of a path $p(u,v)$ between nodes $u$ and $v$. Also, for every node $v$ we define $U_v$ equal to the number of hops used to reach $v$ from root and at first the $U_{root}$ is equaled to zero. 
 We define variable $itr_v$, the time when node $v$ is added to the set $G$.
First we set variable $itr_{root}$ for the root node to be zero, and the first basic node (open facility) that is added to the set $G$ after the root will have the $itr=1$. The algorithm consists of two phases. In phase one (Algorithm 2), we compute $U$ values by generalization of Prim algorithm (see \cite{20}) and in second phase (Algorithm 3) by inspiring Kruskal algorithm idea we try to construct the HCST among all open facilities. 
In phase 2, basic nodes are arranged in descending order due to the amount of their $itr$  and then they would be added to the tree with idea of Kruskal algorithm.
%

\bigskip
\begin{algorithm}[H]
\SetKwInOut{Input}{input}\SetKwInOut{Output}{output}
\Input{An undirected graph.}
\Output{U and itr values.}
{\bf Step 1}. {\bf Initialization} $G=(\{root\},\emptyset)$,

\While{$Q\not\subseteq S$}{

{\bf Step 2}. {\bf Find} nodes $u^*\in V_G$ and $v^*\in V_T$ and path P(u*,v*), where 

$U_v^* = U_u^* +|P(u^*,v^*)|\leq H$,

$d_{u^* v^*}=minÃ¢ÂÂ¡\{d_{uv} |u\leq V_G ,v^*\in V_T\}$;

{\bf Step 3}. {\bf Add} the nodes and edges of path $P(u^*,v^*)$ to G,

{\bf update} U values of the nodes of path $P(u^*,v^*)$,

{\bf Save} $itr[Basic node\hspace{0.2cm} v \in P(u^*,v^* )]$.
}
\caption{NRBI (Phase 1)}
\end{algorithm}
\begin{algorithm}[H]
\SetKwInOut{Input}{input}\SetKwInOut{Output}{output}
\Input{$U_i$ and itr. }
\Output{Construction of HCST.}
{\bf Initialization} $Tree=\{\emptyset\}$

\For{Maximum $Itr_v$ to 1}{

{\bf Find} $P(v,u^* )$ where 

$u^*\in Tree$,

$U_u^*+|P(u^*, v )|\leq U_v$;

\If {$Cost(P(v,u^* ))<Cost(P(v,u^* )\in G)$}

{{\bf Add} all nodes and edges of $P(v,u^* )$ to $Tree$.}

\Else{ {\bf Add} all nodes and edges of $P(v,u^* )\in G$ to $Tree$.}
}
\caption{NRBI (Phase 2)}
\end{algorithm}
\bigskip

As in the first part of the algorithm all $U$ values were calculated and with this precondition that every node can only be connect to the node with lower $U$, in the second part we are sure that the result is tree. The time complexity of this algorithm at the first part is equal to $O(ELogE)$. In the second part in the tree construction, all nodes are connected to each other in the forest, since when every basic node added to the set is compared to the all Steiner and basic nodes in the $Tree$, the running time is $O(QV)$.

For every customer $j\in D$, we find the cheapest possible assignment to facility from Tree.

We finally close the facilities that are part of the Steiner tree, but they are not used at all. In fact no customer is connected to them. Hence, both costs of opening and its connection to the root would be subtracted from the total cost of $y$.
\subsection{Parameters initialization:}

As explained in previous section, we have three main parameters in this algorithm: $HMCR, bw$, and $PR$. There is no rule to select $HMCR$ and $PR$ values and we can choose any value depending on the type of the problem, but through the experience usually it is better to choose values that the probability of selecting a new random element is less than 0.05 and probability of modified harmony memory elements is less than 0.15.
\\ How to figure out $PAR$:

Due to the type of this problem, where the vector variables can be only zero or one, if the variables have changed slightly from the previous value, the closest value is contrast to it $(0\rightarrow 1,1\rightarrow 0)$. By inverting the variables we will become far from optimal solution. So, in the harmony search algorithm for HCConFL problem, when  $PR$ is zero, then $bw$ can be any value and it wont have effect on the algorithm.
\\ How to figure out $HMCR$:

As the algorithm progresses, the diversification should be decreased. We use a good strategy at the beginning of the algorithm. Since the diversification of the search should be high at this stage and the algorithm searches the entire space, we choose $HMCR$ parameter to be 0.96 ( we arrive to it by experiments) and when the algorithm progresses, we converge this parameter to 1 to reduce the randomized part of the algorithm. 
Since solution vector for HCConFL problem, can only include two values, zero or one, when choosing random values for variables in the solution vector we have the same trouble that was explained in the $PAR$ selection, and if we want to select randomly, the both probability of opening and closing is 50$\%$, that somewhat the process will be slower to achieve the optimal solution. 
To solve this problem, rather than a random selection with equal probability, we can use different probabilities for selecting each of the facility. In other words, for each facility $x_i$   the probability $p_i$  at the beginning of the algorithm is defined greedy with a formula based on the cost of opening the facilities and the number of customers and etc. 
In other words, if we choose a random value for $x_i$  that has $p_i$ equal to 90$\%$, it means that the facility is opened with probability 90$\%$ and it will be closed 10$\%$. In this case, the discrete space including only 0 and 1 is changed to a discrete space containing quantities of 1 to 100 and clearly it will influence on the quality of the solution. 
For example, if the random variable value generates 0.4 before considering this greedy approach for both facility $x_i$ and $x_j$, this means zero. However, if the value $p_i$ is 0.5 and $p_j$ is 0.3, then the 0.4 for the first facility means opening and for the second means closing. This probability will change in progression of algorithm when every new vector is added to the harmony memory.
\subsection{Filling the harmony memory and $HMS$:}

In order to initialize the harmony memory, the algorithm generates $HMS$ random vectors and puts them in the harmony memory. If $HMS$ is very large, the quality of vectors, which saved in harmony memory will be decreased. If it is small, we will lose some solution vectors because the new solution is built based on harmony memory. Also, the $HMS$ is related to the number of iterations. The number of iterations has to increase to find the appropriate solution. For HCConFL problem, after considering so many values, we set $HMS$ equal to 50.
\subsection{Improvising a new harmony:}

 According to the three rules mentioned in Section 3, a new harmony is improvised. The $HMCR, PR$ and harmony memory are defined according to the type of the problem. Once we initialize the harmony memory, we start to choose the first element for a new harmony. A random number is generated between 0 and 1. If this number is more than HMCR, the new element is produced randomly; otherwise the algorithm selects the element from the first column of harmony memory randomly according to the greedy algorithm that was proposed.
  We must always be careful of harmony memory distinct vectors and try to avoid adding duplicate vector. The reason is obvious. By adding duplicate vector the quality of the new generated solutions will be better temporarily, but finally it causes the local optimum trap.
\subsection{Algorithm termination condition:}

As we have mentioned in harmony algorithm, the termination condition can be applied to a wide variety of options. In our implementation, the termination happens when there is no improvement in the 1000 last iteration in solution of the harmony memory. Therefore, the algorithm concludes that the solution will not be better and then will be stopped.

\section{Greedy Harmony search algorithm for the HCConFL problem}
\label{}
In this section, we present a new greedy algorithm and combine it with harmony search algorithm and try to solve the problem in a better time. If the number of facilities increases, then the search space will increase exponentially. Due to the optimal solution of this problem, we have seen the maximum number of open facilities in every solution is at most 5 or 6 and the solution with for example 30 open facilities do not exist. Therefore, we tried to limit the number of open facilities in vector space. With doing this, first we have reduced the search space that it will cause rising the convergence, therefore decreasing time of the algorithm. So, we introduce {\it maxOpenFacility}, which is maximum number of open facilities in the solutions. Now we need an appropriate algorithm to reduce the number of open facilities in the solution vector. In other words, during filling the harmony memory, if the number of open facilities in the vector is more than {\it maxOpenFacility}, we will try to use an algorithm to close number of facilities, till the number of open facilities arrives to {\it maxOpenFacility}. It is important the algorithm be accurate. In this algorithm we consider what happens when an open facility is closed. By closing the facility, the facility opening cost and the cost of connecting it to the Steiner tree will be subtracted from the overall cost. Besides, all customers that connect to this facility will be free and the connection costs of all will be subtracted from the overall cost.  But, these customers have to connect to the second closest facility and these new costs should be added to cost of the solution. We continue this up to the number of open facilities be equal to {\it maxOpenFacility}. The steps of the algorithm are presented in Algorithm 4. 

\bigskip
\begin{algorithm}[H]
\While{(numOfFacilities < maxOpenFacility) }{
    {\bf Step 0.}     
              \For{all facilities}{
                               firstNearest = secondNearest=0.
}
    {\bf Step 1}. 
                   \For{all customers}{
                           nearstFacility = nearstfacility that is opened;

                           firstNearest[nearestFacility]+=firstNearest[c]=cost Of[c][nearestFacility];

                           secondNearest[nearestFacility]+=secondNearest[c]=cost Of[c][secondNearestFacility];
                    }
   {\bf Step 2}.
                 \For{all facilities}{
                          costOfClosing[f] = secondNearest[f]- firstNearest[f]- costOfOpening[f]- costOfPath[root][f];
                 }
  {\bf Step 3}.
         To optimize the solution, we cannot still have Steiner tree, so we use cost of the path from the root to the facility as the insurance of facility connection cost to the Steiner tree.
}
 \caption{Greedy algorithm for modifying harmony search to solve HCConFL}
\end{algorithm}
\bigskip

For every customer and facility, we use two variables: firstNearest and secondNearest.

This greedy algorithm can be implemented in time complexity of
\\ $O(numOfFacilities\times numOfCustomers\timesÂ \log(numOfFacilities))$. Since the time is less than time of constructing Steiner tree and calculating the cost function, no extra time will be consumed. While the number of iterations of the algorithm is reduced greatly, then we can see the time to reach a solution considerably decreases.

The harmony parameters are selected as we described in Section 4, but here $HMS$ is considered to be larger and in our implementation it is 150, although before this parameter for solving HCConFL problem was around 50 just for keeping very good solution in harmony memory.
 After this optimization, harmony memory can consider the large number of solution vectors that cause much less iterations to reach the optimal solution. In the implementation, the condition is that if there is no improvement in the 1000 previous iterations, the algorithm will terminate.

\section{Greedy harmony search algorithm with local search}
\label{}
      In Section 5, a greedy algorithm was introduced to supplement and modify the harmony search algorithm for solving HCConFL problem. The algorithm reduces the search space by deleting some facilities. The interesting point of the algorithm in limiting the search space is that there is no need to a heuristic or an approximation algorithm to achieve the optimal solution in reasonable time. 
By combining the greedy algorithm with a local search algorithm, we can provide a search space completely and the optimal solution can be achieved with same quality and even better as heuristic algorithm.
To do this, first we generate solutions around 1000 to 2000 random vectors and then with the greedy algorithm, we improve the vectors that the number of the open facilities is limited around 17 or 18. Now we count the number of opening for every facility and then we choose 18 facilities that they opened more than other. 
We select 18 of the best facilities, so we consider all with the 18 selected facilities in the $2^{17}$ and select the best solution as the final solution.

\section{Computational Experiments}
\label{}
We consider classes of OR-library benchmarks are given in (UFLib) \footnote{http://www.mpi-inf.mpg.de/ departments/d1/projects/benchmarks/UFLP.\\
http://people.brunel.ac.uk/mastjjb/jeb/orlib/steininfo.html} to assess our work. 
Instances were merged from UFLP instances with STP instances, to generate ConFL input graphs in the following way: first $|F|$ nodes of STP instances are selected as the root. The number of facilities, the number of customers, opening costs and assignment costs are provided in UFLP files. 
STP files provide edge-costs and additional Steiner nodes. $mp\{1,2\}$ and $mq\{1,2\}$ instances have been proposed by \cite{12}. They are designed to be similar to UFLP real-world problems and have a large number of near-optimal solutions. 
We took two representatives of the classes $mp$ and $mq$ of sizes $200\times 200$ and $300\times 300$, respectively. Instances $\{C, D\}n$, $n\in\{5,10,15,20\}$ were chosen from the OR-Library as representatives of medium size instances for the STP.

All algorithms for the HCConFLP are coded in C++ and run on an Intel QuadCore, 2.4 Ghz machine with 8 Ghz RAM\footnote{https://github.com/Farzaneh9696/HC-facility-location}.

In Tables 1 to 4 the notations TS, HS, and GHS are used in tables as tabu Search (\cite{0}), Harmony Search and Greedy Harmony Search respectively. The first column shows the name of the instance, column $Obj$ provides the value of optimal objective function; in column CPU Time we present needed time (in seconds) to solve the instances. Tabu search algorithm was proposed by \cite{tabu}. 

Tables show the best performing algorithm is GHS, which solves all the instances to optimality for $H\in \{3,5,7,10\}$. The average running time over all 32 instances for GHS increases from 1.45 seconds (H=3) to 2.197 seconds (H=10). 
For HS increases from 12.77 seconds to 70.43 seconds and for TS, this amount increases from 101.172 to 213.51. We also observe that the complexity of the algorithm increases, and its performance slow down with the increasing size of the assignment graphs and the increasing size of the core graph.

\begin{table}[H]
\begin{center}
\begin{adjustbox}{width=\textwidth,totalheight=\textheight,keepaspectratio}
\input{table1.tex}
\end{adjustbox}
\end{center}
\caption{Comparison of the HS Heuristic with the TS on Large-Scale Instances with $Hop=3$}
\end{table}
\begin{table}
\begin{center}
\begin{adjustbox}{width=\textwidth,totalheight=\textheight,keepaspectratio}
\input{table2.tex}
\end{adjustbox}
\end{center}
\caption{Comparison of the HS Heuristic with the TS on Large-Scale Instances with $Hop=5$}
\end{table}
\begin{table}
\begin{center}
\begin{adjustbox}{width=\textwidth,totalheight=\textheight,keepaspectratio}
\input{table3.tex}
\end{adjustbox}
\end{center}
\caption{Comparison of the HS Heuristic with the TS on Large-Scale Instances with $Hop=7$}
\end{table}
\begin{table}
\begin{center}
\begin{adjustbox}{width=\textwidth,totalheight=\textheight,keepaspectratio}
\input{table4.tex}
\end{adjustbox}
\end{center}
\caption{Comparison of the HS Heuristic with the TS on Large-Scale Instances with $Hop=10$}
\end{table}
\section{Conclusion}
\label{}
In this paper, a greedy harmony search algorithm is utilized for solving HCConFL problem. We have used new greedy algorithm in objective function to find HCST for open facilities. The pre-defined parameters for harmony search are attached, and a variable $HMCR$ is utilized.
 Because of the type of the problem, the variable $PR$ and subsequently $bw$ was ignored. This new version of harmony search is found to be an efficient and robust algorithm for HCConFL. Results of solving HCConFL problem on OR Library instances showed that the modified version is much faster and more efficient. We also presented the results of tabu search for solving HCConFL problem to evaluate our results of harmony search.
\bibliographystyle{model5-names}

\end{document}

%% file: table1.tex
\begin{tabular}{cc c |c c|c c }

\hline
&\multicolumn{2}{c}{TS}  &	\multicolumn{2}{c}{HS}&\multicolumn{2}{c}{GHS}\\ 
\hline	
	&Obj	 &CPUTime	&Obj 	&CPUTime 	&Obj 	&CPUTime	 \\ 

\cline{2-7}

C5mp1&	3188.66&	2.37&	3188.66& 	0.68&	3188.66& 	0.31\\
C5mq1&	4904.25& 	6.51&	4904.25& 	2.07&	4904.25& 	1.25\\
C10mp1& 	3034.25& 	17.12&	3032.99&	 3.41& 3032.99&	1.07\\
C10mq1&	 4512.20&	69.95&	4512.20&	9.58&	4512.20& 	2.34\\
C15mp1&	 2814.03&	106.83&	2814.03&	8.12&	2814.03&	1.03\\
C15mq1&	 4576.70&	357.3&	4576.70&	12.58&	4505.18&	 2.59\\
C20mp1&	 2767.10&	123.65&	2762.97&	9.73&	2762.97&	1.02\\
C20mq1&	 4413.11&	369.98&	4413.11&	13.67&	4413.11&	2.14\\
D5mp1&	3221.18&	 2.36&	3221.18&	 0.36&	3221.18& 	0.45\\
D5mq1&	4787.95&	 10.62&	4787.95&	 4.34&	4548.37&	 1.69\\
D10mp1&	3126.20&	 22.95&	3126.22& 	2.09&	3126.22& 	0.77\\
D10mq1&	4881.63&	 87.51&	4881.63& 	9.80&	4441.01& 	2.08\\
D15mp1&	2896.70&	 6.1&	2896.70& 	4.15&	2896.70	&0.99\\
D15mq1&	4619.17&	 14.7&	4619.17& 	17.75	&4234.28&	2.25\\
D20mp1&	2761.97&	 101.37&	2761.97& 	23.85	&2761.97&	0.81\\
D20mq1&	4418.11&	 360.92&	4418.11&	58.05	&4180.58&	2.10\\
C5mp2&	3411.90& 	2.42&	 3411.90&	1.96&	3321.18	&0.29\\
C5mq2&	4548.37&	6.62&	4548.37&	3.92&	4548.37	&1.39\\
C10mp2&	 3263.02&	17.17&	3263.02& 	8.45&	3126.22& 	0.92\\
C10mq2&	 4441.01&	70.77&	4441.01& 	12.66&	4441.01& 	2.03\\
C15mp2&	 3192.35&	107.25&	3183.64& 	28.85&	2896.70& 	0.90\\
C15mq2&	 4221.58&	358.76&	4221.58&	 11.82&	4221.58&	 2.97\\
C20mp2&	 3151.61&	123.58&	3137.35&	 13.12&	2761.97&	 0.80\\
C20mq2&	 4180.58&	430.09& 	4180.58&	 20.81&	4180.58&	2.80\\
D5mp2&	3386.29&	 2.37&	3386.29&	15.35&	3386.00&	0.49\\
D5mq2&	4813.21&	10.66&	4813.21& 	10.26&	4813.21& 	2.05\\
D10mp2&	3290.83& 	3.63&	3290.83& 	3.43&	3290.83& 	0.89\\
D10mq2&	4610.94& 	13.01&	4610.94& 	12.70&	4610.94& 	2.51\\
D15mp2&	3196.55& 	6.16&	3196.55& 	23.42&	3196.55& 	0.91\\
D15mq2&	4289.06& 	14.74&	4289.06& 	58.13&	4289.06& 	1.89\\
D20mp2&	3143.35& 	70&	3143.35& 	28.83&	3143.35& 	0.86\\
D20mq2& 4188.58&	343.58&	4188.58& 	61.83&	4188.58& 	2.09\\
\hline
\end{tabular}

%% file: table2.tex
\begin{tabular}{cc c |c c|c c }
\hline
&\multicolumn{2}{c}{TS}  &	\multicolumn{2}{c}{HS}&\multicolumn{2}{c}{GHS}\\ 
\hline	
	&Obj	 &CPUTime	&Obj 	&CPUTime 	&Obj 	&CPUTime	 \\ 
\cline{2-7}
C5mp1&	3130.49&	12.58&		3130.49&	2.39&			3130.49&	1.01\\
C5mq1&	4753.89&	44.87&		4753.89&	9.61&			4753.89&	2.45\\
C10mp1&	2796.08&	115.16&		2796.08&	15.74&		2796.08&	1.01\\
C10mq1&	4463.03&	390.66&		4463.03&	39.88&		4463.03&	2.05\\
C15mp1&	2780.97&	124.18&		2778.08&	15.71&		2778.08&	1.40\\
C15mq1&	4435.11&	432.22&		4435.11&	41.15&		4435.11&	2.00\\
C20mp1&	2757.97&	125.54&		2757.97&	18.35&		2757.97&	1.00\\
C20mq1&	4412.11&	419.7&		4426.34&	37.98&		4412.11&	1.21\\
D5mp1&	3087.59&	59.59&		3087.59&	2.91	&		3087.59&	1.33\\
D5mq1&	4732.44&	151.3&		4732.44&	6.05	&		4548.37&	2.05\\
D10mp1&	2893.08&	93.16&		2893.08&	10.37&		2893.08&	1.05\\
D10mq1&	4583.59&	362.25&		4583.59&	94.07&		4583.59&	2.02\\
D15mp1&	2780.97&	73.85&		2780.97&	10.62&		2780.97&	1.03\\
D15mq1&	4452.11&	245.75&		4452.11&	37.41&		4234.28&	2.32\\
D20mp1&	2760.97&	100.68&		2760.97&	39.6	&		2760.97&	1.02\\
D20mq1&	4414.11&	353.26&		4414.11&	91.45&		4180.58&	2.18\\
C5mp2&	3287.12&	76.03&		3287.12&	2.37	&		3287.12&	3.17\\
C5mq2&	4439.18&	173.48&		4439.18&	8.64	&		4439.18&	2.09\\
C10mp2&	3174.09&	115.75&		3174.09&	11.69&		3126.22&	1.07\\
C10mq2&	4266.58&	392.14&		4266.58&	38.84&		4266.58&	2.55\\
C15mp2&	3156.35&	125.92&		3156.35&	11.53&		2896.70&	1.04\\
C15mq2&	4198.58&	427.74&		4198.58&	32.81&		4198.58&	2.11\\
C20mp2&	3137.35&	125.21&		3137.35&	14.8	&		2761.97&	1.35\\
C20mq2&	4180.58&	430.36&		4180.58&	15.55&		4180.58&	2.53\\
D5mp2&	3305.53&	14.18&		3305.53&	19.55&		3305.53&	3.75\\
D5mq2&	4362.96&	52.29&		4362.96&	24.08&		4326.96&	2.07\\
D10mp2&	3209.2&	89.5&			3209.20&	23.27&		3209.20&	1.27\\
D10mq2&	4279.47&	326.58&		4279.47&	106.29&		4279.47&	2.55\\
D15mp2&	3154.35&	76.38&		3154.35&	43.95&		3154.35&	0.96\\
D15mq2&	4212.58&	90&			4212.58&	105.28&		4212.58&	2.62\\
D20mp2&	3142.35&	101.05&		3142.35&	35.94&		3142.35&	0.90\\
D20mq2&	4184.58&	343.58&		4184.58&	96.95&		4184.58&	3.57\\
\hline
\end{tabular}

%% file: table3.tex
\begin{tabular}{cc c |c c|c c }
\hline
&\multicolumn{2}{c}{TS}  &	\multicolumn{2}{c}{HS}&\multicolumn{2}{c}{GHS}\\ 
\hline	
	&Obj	 &CPUTime	&Obj 	&CPUTime 	&Obj 	&CPUTime	 \\ 
\cline{2-7}

C5mp1&	2870.89&	74.88&		2870.89&	9.37	&		2870.89&	1.61\\
C5mq1&	4543.16&	263.06&		4543.16&	34.58&		4543.16&	2.50\\
C10mp1&	2796.1&	123.04&		2791.08&	15.89&		2791.08&	1.04\\
C10mq1&	4452.11&	423.87&		4452.11&	23.87&		4452.11&	2.06\\
C15mp1&	2775.97&	125.18&		2775.97&	14.52&		2775.97&	1.04\\
C15mq1&	4424.11&	434.29&		4424.11&	31.17&		4424.11&	2.13\\
C20mp1&	2757.97&	127.06&		2757.97&	15.83&		2757.97&	1.03\\
C20mq1&	4412.11&	433.95&		4412.11&	33.89&		4412.11&	1.78\\
D5mp1&	2894.74&	60.57&		2894.74&	8.33	&		2894.74&	1.13\\
D5mq1&	4514.03&	217.5&		4514.03&	22.08&		4514.03&	2.91\\
D10mp1&	2820.97&	202.97&		2815.08&	5. 30&		2815.08&	1.09\\
D10mq1&	4472.11&	613.15&		4472.11&	13.60&		4441.01&	2.17\\
D15mp1&	2775.97&	99.41&		2775.97&	1.46	&		2775.97&	1.15\\
D15mq1&	4439.11&	336.91&		4439.11&	40.33&		4234.28&	2.34\\
D20mp1&	2759.97&	107.66&		2759.97&	32.69&		2759.97&	1.08\\
D20mq1&	4413.11&	352.47&		4413.11&	105.35&		4180.58&	2.01\\
C5mp2&	3223.64&	76.03&		3223.64&	12.11&		3223.64&	1.10\\
C5mq2&	4354.77&	266.84&		4354.77&	35.89&		4354.77&	2.30\\
C10mp2&	3174.90&	124.74&		3172.95&	12.03&		3126.22&	1.78\\
C10mq2&	4249.58&	425.33&		4249.58&	42.91&		4249.58&	2.92\\
C15mp2&	3156.35&	122.5&		3156.35&	12.49&		2896.70&	1.05\\
C15mq2&	4196.58&	433.65&		4196.58&	34.34&		4196.58&	2.14\\
C20mp2&	3137.35&	125.74&		3137.35&	15.15&		2761.97&	2.93\\
C20mq2&	4180.58&	435.82&		4180.58&	15.55&		4180.58&	2.12\\
D5mp2&	3228.64&	61.64&		3228.64&	32.40&		3228.64&	3.11\\
D5mq2&	4342.68&	228.84&		4342.68&	81.70&		4326.58&	2.59\\
D10mp2&	3181.64&	32.99&		3181.64&	44.40&		3181.64&	2.57\\
D10mq2&	4238.58&	109.29&		4238.58&	174.82&		4236.58&	3.15\\
D15mp2&	3154.34&	101.10&		3153.35&	38.58&		3153.35&	1.08\\
D15mq2&	4207.58&	337.47&		4207.58&	137.99&		4207.58&	2.82\\
D20mp2&	3142.35&	102.49&		3142.35&	52.15&		3142.35&	1.00\\
D20mq2&	4184.58&	341.08&		4184.58&	133.23&		4184.58&	3.78\\
\hline
\end{tabular}

%% file: table4.tex
\begin{tabular}{cc c |c c|c c }
\hline
&\multicolumn{2}{c}{TS}  &	\multicolumn{2}{c}{HS}&\multicolumn{2}{c}{GHS}\\ 
\hline	
	&Obj	 &CPUTime	&Obj 	&CPUTime 	&Obj 	&CPUTime	 \\ 
\cline{2-7}

C5mp1&	2856.98&	126.24&		2856.98&	15.56&		2856.97&	1.64\\
C5mq1&	4470.34&	423.71&		4470.34&	49.48&		4470.34&	2.58\\
C10mp1&	2793.97&	126.35&		2791.08&	14.66&		2791.08&	1.30\\
C10mq1&	4452.11&	428.12&		4452.11&	31.84&		4452.11&	2.35\\
C15mp1&	2772.97&	124.21&		2772.97&	15.19&		2772.08&	1.08\\
C15mq1&	4424.11&	424.89&		4424.11&	39.51&		4424.11&	2.20\\
C20mp1&	2758.97&	125.97&		2758.97&	16.96&		2757.97&	1.08\\
C20mq1&	4412.11&	428.61&		4412.11&	38.2	&		4412.11&	2.90\\
D5mp1&	2847.16&	104.96&		2846.08&	18.23&		2846.08&	1.13\\
D5mq1&	4514.03&	340.51&		4514.03&	29.35&		4514.03&	2.91\\
D10mp1&	2818.97&	31.98&		2810.08&	5.65&			2810.08&	1.12\\
D10mq1&	4463.03&	103.42&		4463.03&	18.25&		4441.01&	2.24\\
D15mp1&	2773.97&	100.98&		2759.97&	16.89&		2759.97&	1.26\\
D15mq1&	4437.11&	338.78&		4413.11&	140.67&		4234.28&	2.73\\
D20mp1&	2759.97&	109.26&		3203.64&	39.8	&		2759.97&	1.23\\
D20mq1&	4413.11&	348.04&		4312.39&	108.53&		4180.58&	2.01\\
C5mp2&	3222.35&	124.74&		3172.95&	14.04&		3172.64&	1.10\\
C5mq2&	4280.58&	429.62&		4312.58&	54.09&		4280.58&	3.62\\
C10mp2&	3174.64&	124.97&		3172.95&	16.89&		3126.22&	2.04\\
C10mq2&	4227.58&	425.33&		4227.58&	43.85&		4227.58&	3.22\\
C15mp2&	3156.35&	123.47&		3156.35&	16.52&		2896.70&	1.21\\
C15mq2&	4196.58&	439.5&		4196.58&	47.49&		4196.58&	2.72\\
C20mp2&	3137.35&	125.82&		3144.64&	16.37&		2761.97&	3.30\\
C20mq2&	4180.58&	430.62&		4180.58&	51.1	&		4180.58&	2.96\\
D5mp2&	3215.64&	106.11&		3215.64&	54.91&		3211.64&	3.21\\
D5mq2&	4271.58&	353.88&		4271.58&	177.07&		4271.58&	2.84\\
D10mp2&	3181.64&	33.46&		3181.64&	51.34&		3181.64&	2.98\\
D10mq2&	4237.58&	109.09&		4237.58&	187.14&		4236.58&	3.19\\
D15mp2&	3153.35&	101.37&		3153.35&	47.65&		3153.35&	1.46\\
D15mq2&	4202.58&	340.25&		4202.58&	200.66&		4202.58&	3.08\\
D20mp2&	3142.35&	102.09&		3142.35&	219.4&		3142.35&	1.04\\
D20mq2&	4184.58&	341.08&		4184.58&	187.9&		4184.58&	3.90\\
\hline
\end{tabular}